\documentclass[12pt, a4paper]{article}

\usepackage[a4paper, margin=1in]{geometry}
\usepackage[T1]{fontenc}
\usepackage[utf8]{inputenc}
\usepackage{lmodern}
\usepackage{csquotes}

\usepackage[english]{babel}
\usepackage[authordate, backend=biber, date=year, url=false, isbn=false]{biblatex-chicago}
\bibliography{references}
\usepackage{enotez}
\setenotez{
    totoc=section,
    backref
}

\usepackage{authblk}
\usepackage{graphicx}
\usepackage{dcolumn}
\usepackage{verbatim}
\usepackage{enumitem}

\usepackage{url}
\usepackage{hyperref}
\usepackage[hyphenbreaks]{breakurl}

\usepackage{xcolor, soul}
\usepackage[mathlines]{lineno}
\modulolinenumbers[5]

\usepackage{amsmath,amssymb}
\usepackage{bm}
\usepackage{bbm}
\usepackage{dsfont}
\usepackage{braket}
\usepackage{slashed}
\usepackage{tensor}
\usepackage[separate-uncertainty = true]{siunitx}
\usepackage{tabularx}
\usepackage{tikz-cd}

\numberwithin{equation}{section}

\newcommand{\leftrightarrowoverset}[1]{\overset{\text{\tiny$\leftrightarrow$}}{#1}}

\DeclareCiteCommand{\citeyear}
    {}
    {\bibhyperref{\printdate}}
    {\addcomma\space}
    {}

\usepackage{acro}

\DeclareAcronym{qft}{
    short = QFT,
    long = quantum field theory
}
\DeclareAcronym{rqt}{
    short = RQT,
    long = relativistic quantum theory
}
\DeclareAcronym{rqft}{
    short = RQFT,
    long = relativistic quantum field theory
}
\DeclareAcronym{aqft}{
    short = AQFT,
    long = algebraic quantum field theory
}
\DeclareAcronym{rqm}{
    short = RQM,
    long = relativistic quantum mechanics
}
\DeclareAcronym{nrqm}{
    short = NRQM,
    long = non-relativistic quantum mechanics
}
\DeclareAcronym{povm}{
    short = POVM,
    long = positive-operator valued measure
}
\DeclareAcronym{nw}{
    short = NW,
    long = Newton-Wigner
}
\DeclareAcronym{kg}{
    short = KG,
    long = Klein-Gordon 
}
\DeclareAcronym{fw}{
    short = FW,
    long = Foldy-Wouthuysen
}
\DeclareAcronym{cffv}{
    short = CFFV,
    long = Case–Foldy–Feshbach–Villars
}
\DeclareAcronym{ek}{
    short = EK,
    long = Eriksen-Kolsrud
}

\begin{document}

\title{The localization problem: an antinomy between measurability and causal dynamics}
\author{Evan P. G. Gale\thanks{\texttt{e.gale@uq.edu.au}}}
\affil{ARC Centre of Excellence for Engineered Quantum Systems, School of Mathematics and Physics, The University of Queensland, St Lucia, QLD 4072, Australia}

\date{\today}

\maketitle

\begin{abstract}
The localization problem in relativistic quantum theory has persisted for more than seven decades, yet it is largely unknown and continues to perplex even those well-versed in the subject. At the heart of this problem lies a fundamental conflict between localizability and relativistic causality, which can also be construed as part of the broader dichotomy between measurement and unitary dynamics. This article provides a historical review of the localization problem in one-particle relativistic quantum mechanics, clarifying some persistent misconceptions in the literature, and underscoring the antinomy between causal dynamics and localized observables.
\end{abstract}



\tableofcontents

\section{Introduction} \label{sec:introduction}

The concept of localization is as crucial to physics as it is mundane, for it is a part of everyday experience that definite objects occupy finite volumes in space. Yet, the simple question of where things are becomes far more ambiguous once relativity or quantum theory enter the picture:~for the former, there is no unique generalization of the center of mass for a relativistic body; for the latter, there remains the interpretational challenge posed by quantum measurement. Such difficulties are compounded in \ac{rqt}, wherein one must confront a fundamental conflict between localization and relativistic causality, known as the localization problem.


Much like the well-known measurement problem, which pertains to how and why a quantum state collapses to a single outcome, the localization problem poses an equally foundational question, yet has not received anywhere near the same attention. The lack of contemporary discussion stems from an unfortunate historical association, wherein localization has been incorrectly viewed as presupposing a particle ontology---particles being \textit{the} quintessential localized system. This bias against localizability has intensified as~\ac{rqft} emerged as the dominant paradigm 
in the late twentieth century, further marginalizing the role of particles as fundamental entities.
It is unsurprising, then, that recent critics have sought to dispense with particles by rejecting the very basis of localization itself. Most prominently,~\textcite{malament_defense_1996} has challenged the consistency of a particle-based~\ac{rqt} on precisely such grounds. Building on earlier results, Malament developed a no-go theorem establishing the mutual incompatibility between localization and locality, assuming standard postulates of relativity and quantum theory. 
Thus, as per the received view, it is not possible to localize \textit{any} system, and we may rest comfortably knowing that a fundamental treatment of~\ac{rqt} cannot be based on particles.

However, there is something deeply unsettling about this conclusion. To reject a particle ontology is one thing, but to give up localizability entirely is quite another. That is to say, irrespective of one's motivations, the localization problem is independent of any ontological commitments. What matters is whether a physical system \textit{can} be strictly localized, \textit{i.e.}, found with certainty in a finite volume. For without any such notion of localizability---whether for particles, fields, or otherwise---what is the operational meaning of measurement? That is, what is the epistemic bearing of a measurement apparatus that is delocalized over all of space? Similar concerns were raised by~\textcite{saunders_dissolution_1994}, who responded to Malament's theorem on this basis:
\begin{quote}
    But it would be odd to insist on a we-know-not-what in remote (spacelike) vicinities, in order to so much as describe the localized phenomena that we see right before our eyes.~[...]~The reason why we are inclined to insist on~[localization and locality]~is because of the silent partner---the \textit{thing} which is localized in~[some spacetime region]. This \textit{very thing} is not also localized in~[some translated region]. But it does not matter very much that it is a particle; it could as well be a ``system'' or ``charge'', or even a ``laboratory phenomenon'', which is after all what we are interested in~\autocite[p.~93]{saunders_dissolution_1994}.
\end{quote}
In response to such concerns,~\textcite{halvorson_no_2002} contend there are no fundamental difficulties. What we call ``particles'' are just epiphenomena, and while observables are delocalized over spacetime, they can still be approximately defined in some bounded region. In other words, while there is no notion of strict localization, one can still speak of systems weakly localized to a finite volume, albeit always with some infinitesimal probability of finding the system outside this region. For all practical purposes, these weakly localized systems and measurement operations can be approximated as if they were strictly localized.

But for such a radical revision, this answer is too facile. From an epistemic perspective, my credence in things localized here and now should not depend, even in principle, on those things found at the far end of the universe, no matter how unlikely. As I would perhaps controversially argue, the very foundation of empirical observation rests on the commonsense assumption that a measurement apparatus can be placed, with certainty, in a given finite volume. For localization does not derive its significance in isolation---of whether systems are localizable---but rather as it concerns the act of measurement, \textit{viz.}~the relation between observer and observable. 
In giving up localization, one cannot presuppose that observables, as well as states and probability assignments, are still readily and meaningfully defined. It is in this sense that the localization problem remains unresolved, despite persistent efforts since the mid-twentieth century.


The history of the localization problem can be roughly divided into three epochs.\endnote{This account excludes photon localization~\autocite{piron_generalized_1967, bacry_localizability_1988, keller_theory_2005, saari_photon_2012}, and the multiple research programs within the axiomatic and algebraic approaches to~\ac{rqft}. The latter began in the 1960s with the ``strict localization'' of~\textcite{knight_strict_1961} and~\textcite{licht_strict_1963, licht_local_1966}, which was followed by~\textcite{schlieder_remarks_1965}, and the \ac{nw}-adjacent approach of~\textcite{segal_quantum_1964, segal_representations_1967}. A survey of contemporary directions along these lines will be reserved for a future article.} 
Initially, between the 1920--50s, there was intermittent study of relativistic mass-centers in both classical and quantum theory, culminating in the works by~\textcite{pryce_mass-centre_1948} and~\textcite{moller_definition_1949}. The second epoch was initiated by~\textcite{newton_localized_1949}, whose seminal paper provoked great interest in the definition of a relativistic position operator; this comparatively brief period, spanning the 1950-70s, is nicely summarized by~\textcite{kalnay_localization_1971}. The final, contemporary epoch, beginning in the 1970s, has focused on localization in relation to relativistic causality; initiated by~\textcite{schlieder_zum_1971} and~\textcite{hegerfeldt_remark_1974}, this study has developed with further works~\autocite{ali_stochastic_1985, thaller_dirac_1992, fleming_just_1996, malament_defense_1996, hegerfeldt_causality_1998, hegerfeldt_instantaneous_1998, fleming_strange_1999, halvorson_no_2002}, and continues to the present. 

The purpose of this paper is twofold. It provides a historical review of localization in~\ac{rqm}, accessible to physicists and philosophers, emphasizing some lesser-discussed works and the close connection between localization and unitary irreducible representations of the Poincar\'{e} group, and concludes with a summary of my own perspective on the localization problem. This paper does not cover localization in~\ac{rqft}, which has a vast literature of its own, and whose treatment will be reserved for a future article. Throughout this paper, I work in natural units~$\hbar = c = 1$ and adopt the metric convention~$(+ \, - \, - \, -)$, where Greek indices are numbered~$\mu, \nu = \{0, 1, 2, 3\}$ and Latin indices are numbered~$i, j = \{1, 2, 3\}$.

This paper is organized as follows. Sec.~\ref{sec:rel_loc} introduces localization in relativity within both classical and quantum theory, drawing attention to a dichotomy between the orthogonal~\acf{nw} and Lorentz-invariant localization schemes. Sec.~\ref{sec:loc_spin_half} discusses localization in the context of spin-$1/2$ particles, covering the dual unitary representations associated with each localization scheme in connection with the \acf{fw}~transformation. 
Sec.~\ref{sec:causality} then examines the causal implications of localization in relation to key no-go theorems, namely Hegerfeldt's and Malament's theorems, as well as the dynamical impact of the corresponding unitary representations on relativistic propagators. 
Sec.~\ref{sec:loc_problem} summarizes the localization problem in light of the preceding discussion and juxtaposes the two opposing historical positions that have been taken. Finally, in Sec.~\ref{sec:situation}, I offer my own thoughts on the localization problem and situate it more broadly within the foundations of physics, before concluding in Sec.~\ref{sec:conclusion}.

\section{Relativistic localization} \label{sec:rel_loc}


\subsection{Relativistic center of mass}


To introduce our discussion of localization in relativistic mechanics, let us consider the localization of a non-relativistic system. In Newtonian mechanics, the localization of an extended body may be uniquely defined as its mass-center (aka center of mass), which transforms covariantly under the Galilean group. For a body with mass distribution~$\rho(\bm{x})$, its mass-center is given by
\begin{align} \label{eq:cm_coord_cts}
    \bm{x}_{\mathrm{c.m.}} := \frac{1}{M} \int d^3\bm{x} \, \rho(\bm{x}) \, \bm{x} \, ,
\end{align}
where
\begin{align}
    M = \int d^3\bm{x} \, \rho(\bm{x}) \, .
\end{align}
Despite this simplicity, the generalization of a mass-center to relativistic mechanics is by no means straightforward; due to the velocity-dependence of mass-energy, defining the mass-center under the Lorentz group offers much greater freedom.

The Dutch physicist Adriaan D. Fokker~(\citeyear{fokker_dynamische_1927, fokker_relativiteitstheorie_1929}) pioneered the study of the mass-center in classical relativity, although Fokker's work initially received little attention until the exchange with~\textcite{born_mass_1940} in~\textcite{fokker_mass_1940}.\endnote{Unaware of Fokker's work, Born had previously treated the relativistic mass-center in~\ac{rqft}~\autocite{born_quantization_1935}, to which~\textcite{pryce_commuting_1935} responded in the same issue.} The first to provide a comprehensive treatment of the relativistic mass-center in both classical and quantum theory was the British physicist Maurice H. L. Pryce~(\citeyear{pryce_mass-centre_1948}), whose work followed the 1947 Dublin lectures of~\textcite{moller_definition_1949}.

Pryce gave six possible definitions~(a)-(f) generalizing the non-relativistic case, which specify whether the relativistic mass-center is weighted by its energy~$P^0$ or rest mass~$m$ 
and how it depends on the choice of reference frame. Of particular importance are definitions~(c) and~(e)---the former is a Lorentz-covariant definition of the mass-center for a body with energy density~$T^{00}(\bm{x})$, which takes the noncovariant form
\begin{align}
    \bm{x}^{\mathrm{(c)}}_{\mathrm{c.m.}} := \frac{1}{P^0} \int d^3\bm{x} \, T^{00}(\bm{x}) \, \bm{x} \, ,
\end{align}
with
\begin{align}
    P^0 = \int d^3\bm{x} \, T^{00}(\bm{x}) \, .
\end{align}
This is a natural generalization of Eq.~\eqref{eq:cm_coord_cts}, but presents several serious problems. If one tries to define the spin angular momentum~$\bm{S}^{\mathrm{(c)}}$ with respect to this mass-center
\begin{align}
    \bm{S}^{\mathrm{(c)}} := \bm{J} - \bm{x}^{\mathrm{(c)}}_{\mathrm{c.m.}} \times \bm{P} \, ,
\end{align}
where~$\bm{J}$ is the total angular momentum, then~$\bm{S}^{\mathrm{(c)}}$ does not satisfy the usual Poisson bracket relations (formally, an~$\mathfrak{su}(2)$~Lie algebra)
\begin{align}
    \{ S^{\mathrm{(c)}}_i, S^{\mathrm{(c)}}_j \}_{\mathrm{PB}} = \epsilon_{ijk} \left( S^{\mathrm{(c)}}_k - \frac{\bm{S}^{\mathrm{(c)}} \cdot \bm{P}}{(P^0)^2} P_k \right) \neq \epsilon_{ijk} S^{\mathrm{(c)}}_k \, ,
\end{align}
and the Poisson brackets between the mass-center components no longer vanish in general
\begin{align} \label{eq:non_comm_pryce}
    \{ x^{\mathrm{(c)}}_i, x^{\mathrm{(c)}}_j \}_{\mathrm{PB}} = -\frac{1}{(P^0)^2} \epsilon_{ijk} S^{\mathrm{(c)}}_k \neq 0 \, .
\end{align}
This is particularly problematic in quantum theory, where the Poisson brackets become commutators, as the non-commuting components of~$\bm{x}^{\mathrm{(c)}}_{\mathrm{c.m.}}$ would preclude the simultaneous measurement of the mass-center's coordinates. To avoid these issues, one might alternatively require that the Poisson brackets vanish
\begin{align} \label{eq:comm_pryce}
    \{x^{\mathrm{(e)}}_i, x^{\mathrm{(e)}}_j\}_{\mathrm{PB}} = 0 \, , \quad \forall i,j \, ,
\end{align}
which Pryce took as the condition for case~(e), defined as
\begin{align}
    \bm{x}^{\mathrm{(e)}}_{\mathrm{c.m.}} := \bm{x}^{\mathrm{(c)}}_{\mathrm{c.m.}} + \frac{1}{m(m + P^0)} \bm{S}^{\mathrm{(c)}} \times \bm{P} \, .
\end{align}
However, the imposition of commuting components implies the mass-center is no longer covariant and is dependent on the choice of reference frame. In other words, Lorentz covariance and vanishing Poisson brackets are mutually exclusive properties! As will become apparent, this dichotomy also arises for the different possible definitions of localized states and relativistic position operators in~\ac{rqm}.

\subsection{Localized states and relativistic position operators} \label{sec:pos_op}

As shown, a natural way to define an extended body's location is by its mass-center, although other approaches are also possible. An axiomatic treatment was given by~\textcite{newton_localized_1949}, who sought to define localized states in~\ac{rqt} that correspond to the location of an ``elementary system'' or mass-center of an extended body. A more formal treatment was later provided by~\textcite{wightman_localizability_1962}, who gave an alternative axiomatic construction of the \acf{nw}~position operator.

However, like the dichotomy between Pryce's definitions~(c) and~(e) in classical relativity, the same problem occurs in~\ac{rqt}; despite satisfying many natural properties, the \ac{nw}~localization scheme is not Lorentz invariant. In response, \textcite{philips_lorentz_1964} developed an alternative Lorentz-invariant localization scheme, but had to sacrifice strict localization. Bearing these considerations in mind, let us review the \ac{nw}~localization scheme.

\subsubsection{Newton-Wigner localization} \label{sec:nw_loc}

\textcite{newton_localized_1949} sought to define the localized state for an ``elementary system,'' or more formally, an irreducible unitary representation of the Poincar\'{e} group, which were previously classified in a seminal work by~\textcite{wigner_unitary_1939}. \ac{nw}~postulated a set of states~$S_0$  representing a system initially localized at the origin of a coordinate system, and required the following four conditions:
\begin{enumerate}[label=(\roman*)]
    \item \textbf{Linearity:} $S_0$ is a linear set, where superpositions of localized states are also localized.
    \item \textbf{Rotation-reflection invariance:} $S_0$ is invariant under rotations, parity transformations, and time reversal.
    \item \textbf{Orthogonality:} $S_0$ is orthogonal to all spatially translated states.
    \item \textbf{Regularity:} The localized states are mathematically well-behaved---all generators of the Lorentz group are applicable to the localized states.\endnote{I have repeated here the standard presentation of postulate~(iv) as a regularity condition, which is necessary to ensure the localized state is uniquely defined; see~\textcite{galindo_uniqueness_1965} for a formal treatment. This condition can also be given a proper physical interpretation: it postulates the existence of a rest frame, relative to which the localized state is uniquely defined~\autocite{mathews_observables_1962}.} 
\end{enumerate}
Requiring a state~$\ket{\psi} \in S_0$ satisfy these four conditions, \ac{nw}~derived their eponymous localized state and relativistic position operator. Provided that the Hilbert space inner product between two localized states with spin~$\sigma$ is defined to be
\begin{align} \label{eq:canonical_inner-product}
    \braket{\psi | \phi} := \int d^3\bm{p} \, \psi^*(\bm{p}, \sigma) \phi(\bm{p}, \sigma) = \int d^3\bm{x} \, \psi^*(\bm{x}, \sigma) \phi(\bm{x}, \sigma) \, ,
\end{align} 
which is the standard $L^2$-inner product as in \ac{nrqm}, then one finds the \ac{nw}~wavefunction given by the Fourier transform
\begin{align} \label{eq:nw_wf}
    \psi_{\mathrm{NW}}(\bm{x}, \sigma) = \frac{1}{(2 \pi)^3} \int d^3\bm{p} \, \widetilde{\psi}_{\mathrm{NW}}(\bm{p}, \sigma) \, e^{-i \bm{p} \cdot \bm{x}} \, ,
\end{align}
which due to the positive definiteness of the $L^2$-inner product, can be interpreted as a probability amplitude for measuring a particle or elementary system at position~$\bm{x}$. Similarly, the position eigenstate is given by
\begin{align} \label{eq:nw_pos}
    \ket{\bm{x}, \sigma}_{\mathrm{NW}} = \frac{1}{(2 \pi)^3} \int d^3\bm{p} \, e^{-i \bm{p} \cdot \bm{x}} \ket{\bm{p}, \sigma} \, ,
\end{align}
where
\begin{align}
    \braket{\bm{p}, \sigma | \bm{p}', \sigma'} = (2 \pi)^3 \, \delta^{(3)}(\bm{p} - \bm{p}') \delta_{\sigma \sigma'} \, .
\end{align}
Following from this construction of localized states, one can define a unique position operator~$\bm{x}_{\mathrm{NW}}$, which is self-adjoint and also preserves positivity of the energy. Moreover, adopting the $L^2$-inner product, this position operator has the standard representations in configuration and momentum space, namely\endnote{\textcite{newton_localized_1949} employ the Lorentz-invariant integration measure~$d^3\bm{p} / p_0$ with energy~$p_0$, which leads to a rather unwieldy form for the \ac{nw}~position operator, given by~$$ [\bm{x}_{\mathrm{NW}}]_p = i \left( \bm{\nabla}_p - \frac{\bm{p}}{2 p_0^2} \right) = p_0^{1/2} \, i \bm{\nabla}_p \, p_0^{-1/2} \, . $$ However, since the \ac{nw}~localization scheme lacks Lorentz invariance, and given the simple form of the position operator with respect to the standard measure~$d^3\bm{p}$, there is little reason to adopt~$d^3\bm{p} / p_0$. See~\textcite{fong_bra-ket_1968} for further discussion of these different conventions.}
\begin{align}
    [\bm{x}_{\mathrm{NW}}]_x = \bm{x} \; \text{ and } \; [\bm{x}_{\mathrm{NW}}]_p = i \bm{\nabla}_p \, .
\end{align}
However, despite all of the above-listed properties, the \ac{nw}~scheme suffers from many key deficiencies---it is not preserved under Lorentz transformations or time translations. As can be seen from conditions~(i)-(iv), the \ac{nw}~scheme is defined only with respect to the Euclidean group, rather than the full Poincar\'e group. Furthermore, as~\textcite{wightman_localizability_1962} points out, the axioms necessarily imply that~$\bm{x}_{\mathrm{NW}}$ has commuting components 
\begin{align} \label{eq:nw_comm}
    [x_{\mathrm{NW}}^i, x_{\mathrm{NW}}^j] = 0 \, , \quad \forall i,j \, ,
\end{align}
which means that the \ac{nw}~scheme is equivalent to Pryce's~(e), as both are exclusively characterized by this requirement. 
Thus, just as for~(e), the \ac{nw}~scheme depends on the choice of hyperplane, \textit{i.e.}, a complete definition of a \ac{nw}~localized state is
\begin{align}
    \ket{(\bm{x}, t), \Sigma} \; \text{ with coordinate } (\bm{x}, t) \text{ \textit{and} hyperplane } \Sigma \, .
\end{align}
This hyperplane dependence has been strongly emphasized by the American physicist Gordon N. Fleming~(\citeyear{fleming_covariant_1965, fleming_manifestly_1966, fleming_hyperplane_1989, fleming_just_1996, fleming_strange_1999}), whose views will be discussed in Sec.~\ref{sec:loyalism}. For now, let us turn to Philips' Lorentz-invariant localization scheme.


\subsubsection{Philips localization} \label{sec:ph_loc}

An alternative set of conditions was given by \textcite{philips_lorentz_1964}, who sought a Lorentz-invariant notion of localization---the implicit motivation being to define an elementary system that remains localized in a given spacetime region, rather than spatial volume. To formalize this scheme, Philips retained the first two conditions of~\ac{nw}, but had to drop (iii)~orthogonality due to its incompatibility with Lorentz invariance. Consequently, the inner product between two Philips states never fully vanishes, even for arbitrarily large displacements in spacetime, so it is somewhat of a misnomer to call this a ``localization'' scheme. For a set~$S_0$ of localized states at the origin, Philips' conditions are:
\begin{enumerate}[label=(\roman*)]
    \item \textbf{Linearity:} $S_0$ is a linear set, where superpositions of localized states are also localized.
    \item \textbf{Rotation-reflection invariance:} $S_0$ is invariant under rotations, parity transformations, and time reversal.
    \item \textbf{Lorentz invariance:} $S_0$ is invariant under Lorentz boosts.
    \item \textbf{Regularity:} Two additional conditions to ensure the localized states are mathematically well-behaved---normalizability and irreducibility.
\end{enumerate}
Given the Lorentz invariance of the localization scheme, one naturally adopts a Lorentz-invariant inner product; for the spin-$0$ case, this corresponds to the \ac{kg}~inner product in configuration space~\autocite{fong_bra-ket_1968}
\begin{align}
    (\psi, \phi) := \int \frac{d^3\bm{p}}{2 p_0} \, \psi^*(\bm{p}) \phi(\bm{p}) = i \int_t d^3\bm{x} \, \psi^*(\bm{x}, t) \leftrightarrowoverset{\partial}_t \phi(\bm{x}, t) \, ,
\end{align}
where~$\leftrightarrowoverset{\partial} := \overset{\rightarrow}{\partial} - \overset{\leftarrow}{\partial}$, and the integral is taken over a spacelike hypersurface at fixed time~$t$. Adopting this definition, Philips obtained the unique Lorentz-covariant position state\endnote{I am skipping over some subtleties here---a conjecture Philips assumed to ensure uniqueness was later shown to be false \autocite{gallardo_lorentz-invariant_1967}. However, these Lorentz-covariant position states are still well-defined, and their uniqueness for spin-$0$ was later proven by \textcite{kalnay_lorentz-invariant_1970}.}
\begin{align} \label{eq:philips_pos}
    \ket{\bm{x}}_{P} = \frac{1}{(2 \pi)^3} \int \frac{d^3\bm{p}}{2 p_0} \, e^{-i \bm{p} \cdot \bm{x}} \ket{\bm{p}} \, ,
\end{align}
where~$p_0 = \sqrt{p^2 + m^2}$ is the energy, and unlike for the \ac{nw}~scheme, the momentum eigenstates have been covariantly normalized\endnote{This normalization is purely conventional and does not impact the dynamics. For spin-$0$ particles, the \ac{nw} and Philips wavefunctions are related by~$\psi_{\mathrm{NW}} = \psi_{P} / \sqrt{p_0}$, which \textit{is} a physically significant rescaling.}
\begin{align}
    \braket{\bm{p} | \bm{p}'} = 2 p_0 \, (2 \pi)^3 \, \delta^{(3)}(\bm{p} - \bm{p}') \, .
\end{align}
Likewise, the Philips wavefunction is also Lorentz covariant, but is no longer given by a Fourier transform due to the introduction of a Lorentz-invariant integration measure
\begin{align} \label{eq:philips_wavefunction}
    \psi_{P}(\bm{x}) = \frac{1}{(2 \pi)^3} \int \frac{d^3\bm{p}}{2 p_0} \, \widetilde{\psi}_{P}(\bm{p}) \, e^{-i \bm{p} \cdot \bm{x}} \, .
\end{align}
However, while in the \ac{nw}~scheme one can interpret its wavefunction as a probability amplitude, \textit{no such interpretation} is possible for the Philips wavefunction. The \ac{kg}~inner product is not positive definite, so it cannot be interpreted as a position probability density---if this were true then one would have negative probabilities!\endnote{For further discussion on issues with the \ac{kg}~inner product, see~\textcite{fleming_strange_1999}.} Regardless, by adopting this inner product, the Philips position operator~$\bm{x}_{P}$ has the standard representations in configuration and momentum space~\autocite{gallardo_philips_1967}
\begin{align}
    [\bm{x}_{\mathrm{P}}]_x = \bm{x} \; \text{ and } \; [\bm{x}_{\mathrm{P}}]_p = i \bm{\nabla}_p \, .
\end{align}
Nevertheless, replacing orthogonality with Lorentz invariance causes multiple problems; not only does non-orthogonality prohibit strict localization, it also problematizes~$\bm{x}_{P}$, which as Philips laments:
\begin{quote}
    The nonorthogonality of the eigendifferentials means that there is no self-adjoint operator (``position operator'') [...]. This constitutes an unfortunate consequence of the decision to drop the orthogonality requirement included in the NW postulates~\autocite[p.~896]{philips_lorentz_1964}.
\end{quote}
Even worse, the Philips position operator is also non-normal~\autocite{gallardo_philips_1967}
\begin{align}
    [\bm{x}_{P}, \bm{x}_{P}^\dagger] \neq 0 \, ,
\end{align}
which casts serious doubt on its status as an observable---though that has not stopped people from trying~\autocite{kalnay_reinterpretation_1967}.


Noting the significance of the \ac{kg}~inner product, one might wonder how the Philips scheme generalizes for higher spin. In fact, Philips considered just the spin-$0$ case. Higher-spin generalizations were treated only later by other authors~\autocite{gallardo_lorentz-invariant_1967, kalnay_lorentz-invariant_1970}, who had great difficulty in extending Philips' approach, primarily due to the regularity conditions~(iv). While I will not discuss the attempts at generalizing the Philips scheme, it is worthwhile to consider the localization and dynamics of spin-$1/2$ particles, which have some notable particularities.

\section{Localization and dynamics of \texorpdfstring{spin-$1/2$}{spin-1/2} particles} \label{sec:loc_spin_half}

The story of Dirac's achievement in obtaining a relativistic wave equation for spin-$1/2$ particles is well-known [see~\textcite{saunders_negative-energy_1991} for a retelling]. Trying to avoid the negative-energy solutions of the \ac{kg}~equation, \textcite{dirac_quantum_1928} sought a wave equation linear in both space and time, whose Hamiltonian squared to the usual relativistic dispersion relation. The resulting Hamiltonian is defined as
\begin{align} \label{eq:dirac_hamiltonian}
    H_{\mathrm{D}} = \bm{\alpha} \cdot \bm{p} + \beta m \, ,
\end{align}
from which one obtains the eponymous Dirac equation in its non-covariant form
\begin{align} \label{eq:dirac_eq}
    i \partial_t \psi(\bm{x}, t) = \left( \bm{\alpha} \cdot \bm{p} + \beta m \right) \psi(\bm{x}, t) \, ,
\end{align}
where~$\psi \in \mathbb{C}^4$ are Dirac spinors. The matrices~$\bm{\alpha}$ and $\beta$ satisfy a Clifford algebra
\begin{align}
    \{ \alpha_i, \alpha_j \} = 2 \delta_{ij} \, , \quad \{ \alpha_i, \beta \} = 0 \, , \quad \alpha_i^2 = \beta^2 = 1 \, ,
\end{align}
which in the standard Dirac basis take the form
\begin{align}
    \alpha_i =
    \begin{pmatrix}
        0        & \sigma_i \\
        \sigma_i & 0
    \end{pmatrix} \, , \quad 
    \beta =
    \begin{pmatrix}
        I & 0 \\
        0 & -I
    \end{pmatrix} \, ,
\end{align}
where~$\sigma_i$ are the Pauli matrices and~$I$ denotes the~$2 \times 2$ identity matrix. Conventionally, one chooses the 
positive-definite inner product~\autocite[p. 7]{thaller_dirac_1992}
\begin{align} \label{eq:dirac_inner_product_canonical}
    \braket{\psi | \phi}_D := \int d^3\bm{x} \, \psi^\dagger(\bm{x}) \phi(\bm{x}) \, ,
\end{align}
where~$\psi^\dagger$ denotes the conjugate transpose. While Dirac had hoped to find a wave equation free of negative-energy solutions, the formulation still yielded a Hamiltonian with an unbounded spectrum. Dirac interpreted these negative-energy solutions in terms of ``hole theory'', which as per standard dogma, paved the way for~\ac{rqft}. Let us examine some lesser known features of the Dirac equation, namely its different unitary representations, and their relation to the localization and dynamics of spin-$1/2$ particles.

\subsection{Foldy-Wouthuysen transformation}



Considering a spin-$1/2$ particle with electric charge~$e$, which is minimally coupled to an electromagnetic field~$A^\mu = (\varphi, \bm{A})$, the Dirac Hamiltonian becomes
\begin{align}
    H_{\mathrm{D}} = \bm{\alpha} \cdot (\bm{p} - e \bm{A}) + \beta m - e \varphi \, .
\end{align}
Since it is not possible to directly take the non-relativistic limit of this Hamiltonian, one traditionally obtains the non-relativistic limit of the Dirac equation~\eqref{eq:dirac_eq} by factoring out the rest energy and neglecting certain rapidly oscillating solutions~\autocite[pp.~168-169]{pauli_general_1980}. In so doing, however, one obtains an imaginary electric dipole moment, which breaks the Hermiticity of the Hamiltonian and spoils the non-relativistic limit.


\textcite{foldy_dirac_1950}~recognized that the ``odd'' operator~$\bm{\alpha}$, which couples the particle and antiparticle components of the Dirac spinor, is responsible for the incorrect non-relativistic limit. Foldy sought to remove this odd operator, and succeeded in deriving a unitary transformation that gave a series expansion in powers of~$1/m$. In the free case, without coupling~$A^\mu$, Wouthuysen showed that this transformation can be expressed by the following unitary involution\endnote{Although widely attributed to~\textcite{foldy_dirac_1950}, this transformation (for free particles) had been previously obtained by~\textcite[p.~71]{pryce_mass-centre_1948}, and the Japanese physicist Smio Tani~(\citeyear{tani_dirac_1949, tani_connection_1951}).}
\begin{align} \label{eq:foldy_wouthuysen}
    U_{\mathrm{FW}} = \exp\left[ \frac{\beta \bm{\alpha} \cdot \bm{p}}{2 |\bm{p}|} \tan^{-1}\left( \frac{|\bm{p}|}{m} \right)\right] = \frac{\beta H_\mathrm{D} + E_{\bm{p}}}{\sqrt{2 E_{\bm{p}} (E_{\bm{p}} + m)}} \, .
\end{align}
This \acf{fw}~transformation diagonalizes the Dirac Hamiltonian~\eqref{eq:dirac_hamiltonian}, so that the new representation involves only ``even'' operators\endnote{An ``even'' operator does not mix the positive- and negative-energy components of the Dirac spinor, while an ``odd'' operator does couple such components.} 
\begin{align} \label{eq:dirac_hamiltonian_canonical}
    H^{\mathrm{(C)}}_{\mathrm{D}} := U_{\mathrm{FW}} H_{\mathrm{D}} U_{\mathrm{FW}}^\dagger = \beta \sqrt{\bm{p}^2 + m^2} \equiv \beta E_{\bm{p}} \, ,
\end{align}
which is called the ``canonical'' representation (denoted~``$C$''); I term the original the ``hyperbolic'' representation (without denotation).\endnote{The ``canonical'' representation~\autocite{foldy_synthesis_1956} goes by many names in the literature, including ``classical''~\autocite{bose_representations_1959}, ``\ac{fw}''~\autocite{rose_uniqueness_1961}, and ``\ac{nw}''~\autocite{costella_foldywouthuysen_1995}. The ``hyperbolic'' representation has no standard name for arbitrary spin, but is conventionally called the ``Dirac'' representation for spin-$1/2$ systems.} 
Since~$\beta$ is an even operator, with positive~$+1$ (negative~$-1$) eigenvalues for (anti)particles, the positive- and negative-energy components of the Dirac spinor are now decoupled. In a posthumously published memoir,~\textcite{foldy_origins_2006} recounts the derivation of their eponymous transformation:
\begin{quote}
    I told my friend Sieg about what I had found in the hope of getting his help in carrying out and checking these and planned future calculations. A day or two later he came beck~[sic]~to me with the result that he could obtain the canonical transformation for a free particle in closed form (rather than as an infinite series) which would completely decouple the upper and lower components of the Dirac wave-function. It contained a square-root of an operator involving the momentum!~[...]~I soon realized~[...]~that here was the key to most of the puzzles about the Dirac equation~\autocite[p.~349]{foldy_origins_2006}.
\end{quote}
The significance of the \ac{fw}~transformation, to which Foldy eludes, bears precisely on different localizations and dynamics of the two unitary representations. These qualities are best demonstrated by considering the Dirac velocity and spin operators.

\subsection{Velocity, spin, and position} 

The same year Dirac derived his eponymous wave equation, the American physicist Gregory Breit~(\citeyear{breit_interpretation_1928}) noticed a peculiar feature. Defining the Dirac position operator~$\bm{x}_{\mathrm{D}}$ to have the standard representations~\autocite[p.~7]{thaller_dirac_1992}
\begin{align}
    [\bm{x}_{\mathrm{D}}]_x = \bm{x} \; \text{ and } \; [\bm{x}_{\mathrm{D}}]_p = i \bm{\nabla}_p \, ,
\end{align}
the velocity operator is found to be (restoring the speed of light~$c$)
\begin{align} \label{eq:velocity_dirac}
    \bm{v}_{\mathrm{D}} := i [H_{\mathrm{D}}, \bm{x}_{\mathrm{D}}] = \bm{\alpha} c \, ,
\end{align}
which has eigenvalues~$\pm c$, even for massive particles! In other words, if the Dirac velocity operator were treated as an observable, then one would always measure the speed of light. Such an interpretation is highly questionable, however, as the different components of the velocity operator do not commute
\begin{align}
    [\alpha_i, \alpha_j] \neq 0 \, ,
\end{align}
implying that the velocity cannot be simultaneously defined in two different directions, and prohibiting the possibility of velocity measurements~\autocite[p.~91]{schweber_introduction_1961}.
Classically, however, one would expect the relativistic velocity operator to have the form
\begin{align}
    \bm{v} = \bm{\nabla}_p H = \frac{\bm{p}}{p_0} \, ,
\end{align}
which follows from the square-root Hamiltonian~$H = \sqrt{\bm{p}^2 + m^2}$. This result is precisely what one obtains from the canonical Hamiltonian~\eqref{eq:dirac_hamiltonian_canonical} together with the \ac{nw}~scheme
\begin{align}
    \bm{v}^{\mathrm{(C)}}_{\mathrm{NW}} := i [H^{\mathrm{(C)}}_{\mathrm{D}}, \bm{x}^{\mathrm{(C)}}_{\mathrm{NW}}] = \beta \frac{\bm{p}}{p_0} \, ,
\end{align}
which \textit{does} have commuting components, so can be sensibly interpreted as an observable. Moreover, as can be seen from this velocity operator, the \ac{nw}~position operator~$\bm{x}_{\mathrm{NW}}$ takes the standard forms in the canonical representation\endnote{\textcite{foldy_dirac_1950} initially called~$\bm{x}^{\mathrm{(C)}}_{\mathrm{NW}}$ the ``mean position operator,'' which~\textcite[p.~580]{foldy_synthesis_1956} only later recognized as the \ac{nw}~position operator.}
\begin{align}
    [\bm{x}^{\mathrm{(C)}}_{\mathrm{NW}}]_x = \bm{x} \; \text{ and } \; [\bm{x}^{\mathrm{(C)}}_{\mathrm{NW}}]_p = i \bm{\nabla}_p \, .
\end{align}
However, while the~\ac{fw}~transformation maps between the canonical and Dirac Hamiltonians, it only indirectly maps between localization schemes
\begin{align}
    \bm{x}^{\mathrm{(C)}}_{\mathrm{D}} = U_{\mathrm{FW}} \bm{x}_{\mathrm{D}} U_{\mathrm{FW}}^\dagger = \bm{x}^{\mathrm{(C)}}_{\mathrm{NW}} - \frac{i \beta \bm{\alpha}}{2 E_{\bm{p}}} + \frac{i \beta (\bm{\alpha} \cdot \bm{p}) \bm{p} - (\bm{\Sigma} \times \bm{p}) p}{2 E_{\bm{p}} (E_{\bm{p}} + m) p} \, ,
\end{align}
where the spin angular momentum is defined as
\begin{align}
    \bm{\Sigma} = \frac{i}{2} (\bm{\alpha} \times \bm{\alpha}) = \begin{pmatrix}
        \bm{\sigma} & 0 \\
        0           & \bm{\sigma}
    \end{pmatrix} \, .
\end{align}
Likewise, defining the relativistic spin operator~$\bm{S}_{\mathrm{D}} = \bm{\Sigma} / 2$, one finds spin and momentum are similarly entangled when passing between representations
\begin{align}
    \bm{S}^{\mathrm{(C)}}_{\mathrm{D}} = U_{\mathrm{FW}} \bm{S}_{\mathrm{D}} U_{\mathrm{FW}}^\dagger = \bm{S}_{\mathrm{D}} + \frac{i \beta (\bm{\alpha} \times \bm{p})}{2 E_{\bm{p}}} - \frac{\bm{p} \times (\bm{\Sigma} \times \bm{p})}{2 E_{\bm{p}} (E_{\bm{p}} + m)}
\end{align}
This entangling between position, spin, and momentum was previously noted by~\textcite{pryce_commuting_1935, pryce_mass-centre_1948}, namely in transforming between the different relativistic mass-centers; for a contemporary treatment of the above relations, see~\textcite{zou_position_2020}. These results imply that localization is inseparable from dynamics, with both depending on the choice of unitary representation. Let us examine this in more detail.

\subsection{Dual representations of the Poincar\'e group} \label{sec:dual_rep}

As seen above, there are two distinct representations for spin-$1/2$ particles: one associated with hyperbolic dynamics, and the other naturally coinciding with the \ac{nw}~localization scheme. In a concise review of the \ac{fw}~transformation, \textcite{costella_foldywouthuysen_1995} summarize the situation as follows:
\begin{quote}
    [T]here are \textit{two} representations of the Dirac equation that are singled out above all others---each having qualities unique to itself---that have a truly direct correspondence with Nature: The Dirac-Pauli [\textit{i.e.}, hyperbolic] representation is unique due to its linearity; it is the representation in which the charged leptons are minimally coupled. The Newton-Wigner [\textit{i.e.}, canonical] representation is unique due to its decoupling of positive- and negative-energy states; it is the representation in which the operators of the theory correspond to their classical counterparts~\autocite[p.~1121]{costella_foldywouthuysen_1995}.
\end{quote}
While the \ac{fw}~transformation was originally applied only to the Dirac equation, it was later generalized for spin-$0$ and spin-$1$ particles by \textcite{case_generalizations_1954}, whose work was followed by~\textcite{foldy_synthesis_1956} and~\textcite{feshbach_elementary_1958}. 
For spin-$0$ particles in the hyperbolic representation, their evolution is described by the \ac{kg}~equation
\begin{align} \label{eq:klein_gordon_eq}
    \left( \Box + m^2 \right) \phi(x) = 0 \, ,
\end{align}
which can be transformed into the canonical form by taking a complex linear combination of~$\phi$ and its conjugate momentum~$\pi = \partial_t \phi$, giving the \ac{nw}~wavefunction 
\begin{align} \label{eq:fw_spin0}
    \psi_{\mathrm{NW}}(\bm{x}, t) = \frac{1}{\sqrt{2}} \left( R^{-1/2} \pi(\bm{x}, t) - i R^{1/2} \phi(\bm{x}, t) \right) \, ,
\end{align}
with
\begin{align}
    R := \sqrt{-\bm{\nabla}^2 + m^2} \, .
\end{align}
Independently of the aforementioned authors, the American physicist Irving E. Segal~(\citeyear{segal_quantum_1964}) also derived this transformation in the algebraic approach to~\ac{rqft}, namely as a mapping of the Cauchy data at fixed time to a space of localized solutions. For some discussion of the Segal approach, see~\textcite{saunders_mathematical_1988, saunders_negative-energy_1991, saunders_locality_1992}.

For arbitrary spin 
the hyperbolic representation can be defined by the Bargmann-Wigner equations~\autocite{bargmann_group_1948}, which are a system of coupled Dirac equations for a multispinor solution. 
Conversely, for the canonical representation~$\mathcal{H}^{\mathrm{(C)}}$, the \ac{nw}~wavefunction for a spin-$s$ particle is given by
\begin{align} \label{eq:canonical_form}
    i \partial_t \psi_{\mathrm{NW}}^{(s)}(\bm{x}, t) = \beta \sqrt{-\bm{\nabla}^2 + m^2} \, \psi_{\mathrm{NW}}^{(s)}(\bm{x}, t) \, ,
\end{align}
where~$\psi_{\mathrm{NW}}^{(s)} \in \mathcal{H}^{\mathrm{(C)}}$ is a reducible representation of the Poincar\'{e} group, which decomposes as the direct sum of two $(2s + 1)$-component wavefunctions
\begin{align}
    \psi_{\mathrm{NW}}^{(s)} =
    \begin{pmatrix}
        \psi_{\mathrm{+}}^{(s)} \\ \psi_{\mathrm{-}}^{(s)}
    \end{pmatrix} \, , \quad \mathcal{H}^{\mathrm{(C)}} = \mathcal{H}^{\mathrm{(C)}}_{+} \oplus \mathcal{H}^{\mathrm{(C)}}_{-} \, ,
\end{align}
while the canonical Hamiltonian is block-diagonalized and similarly decomposes as 
\begin{align}
    H^{\mathrm{(C)}} = H^{\mathrm{(C)}}_{+} \oplus H^{\mathrm{(C)}}_{-} \, .
\end{align}
In other words, the canonical representation decouples the positive- and negative-energy solutions, where the \ac{nw}~wavefunction naturally decomposes into the direct sum of two antiunitarily equivalent irreducible representations~\autocite{foldy_synthesis_1956}. That the canonical representation exists at all, however, is rather peculiar; this state-of-affairs was commented on by~\textcite{saunders_dissolution_1994}:
\begin{quote}
    A further complication is that in the midst of [\ac{rqt}] we have what is usually called the Newton-Wigner representation~[...]~in which fields and states are no longer covariantly described and where the ``local'' self-adjoint quantities (NW-local observables) do not obey microcausality, but only satisfy equal-time commutators with respect to a particular inertial frame. If we pass to this representation, we obtain at a stroke the basic structure of non-relativistic quantum field theory; it is not too hard to descend from that to a many-particle mechanics, and to recover the usual definitions of localization~\autocite[p.~88]{saunders_dissolution_1994}.
\end{quote}
That one can easily reobtain~\ac{nrqm} from the canonical representation is not merely a convenient choice, but is formally required to contract the Poincar\'e to the Galilean group. While the group contraction of~\textcite{inonu_contraction_1953} is widely known, their limiting procedure is defined only over the Lie algebra, which leaves open how to contract the full group structure. As argued by~\textcite{leon_probabilistic_1978}, one requires coincident representations for group contraction:
\begin{quote}
    [T]he definition of a contracted representation implies the existence of a family of unitary transformations between representation spaces of both groups. This family does exist if we take probabilistic amplitudes for representing the Poincar\'{e} group. However, if we take as starting point invariant functions, the contracted states are not well defined in the limit~$c \to \infty$. Thus the choice of probability amplitudes for contracting Poincar\'{e} to Galilei group is by no means arbitrary but inherent to the method~\autocite[pp.~118-119]{leon_probabilistic_1978}.
\end{quote}
Thus, to recover~\ac{nrqm}, one must work in the Hilbert space of $L^2$-functions on the Poincar\'{e} group, which necessitates mapping to the canonical representation. One might think the story ends here. Surprisingly, however, the choice of representation is also relevant for~\ac{nrqm}. 
Considering different proper unitary representations of the Galilean group, \textcite{inonu_representations_1952} found a conflict between localization and states with a definite velocity; this conflict is usually avoided by restricting to projective representations of the Galilean group---a move unavailable for the Poincar\'e group. Having considered the role of unitary representations, let us consider the dichotomy between localization and causal dynamics from another perspective.

\section{Causality and relativistic dynamics} \label{sec:causality}

\subsection{No-go theorems for strict localization}

After initial interest in localized states and relativistic position operators, focus shifted in the 1970s to the causal implications of localization, leading to several no-go theorems between localization and causal dynamics. A seminal paper was given by~\textcite{schlieder_zum_1971}, whose work was later developed by~\textcite{busch_unsharp_1999}. However, the most significant result of this period is Hegerfeldt's theorem~(\citeyear{hegerfeldt_remark_1974}), which was subsequently strengthened over the following decades~\autocite{hegerfeldt_remarks_1980, hegerfeldt_violation_1985, hegerfeldt_causality_1998, hegerfeldt_instantaneous_1998, hegerfeldt_particle_2001}. Motivated by these works, \textcite{malament_defense_1996} developed a similar no-go theorem aiming to rule out the existence of particles in~\ac{rqt}. These no-go theorems were further refined by~\textcite{halvorson_no_2002}, who generalized them for~\acp{povm} in globally hyperbolic spacetimes.

\subsubsection{Hegerfeldt's theorem}

Without assuming a particular localization scheme,~\textcite{hegerfeldt_remark_1974, hegerfeldt_causality_1998, hegerfeldt_instantaneous_1998} established a no-go theorem between localization and causality. For a localized system with state~$\ket{\psi} \in \mathcal{H}$, Hegerfeldt postulated that there exists a self-adjoint operator~$N(V)$ giving the probability of finding the localized system in fixed volume~$V$, such that
\begin{align}
    0 \leq \braket{\psi | N(V) | \psi} \leq 1 \, .
\end{align}
This operator~$N$ can be concretely realized as a projector, such that for a system strictly localized in~$V$, one has
\begin{align}
    N(V) \ket{\psi} = \ket{\psi} ,
\end{align}
while for systems not strictly localized in~$V$, one instead finds
\begin{align}
    N(V) \ket{\psi} = 0 \, .
\end{align}
Adopting this definition for localization, Hegerfeldt made two further assumptions restricting the possible dynamics of the system. Specifically, Hegerfeldt assumed: 
\begin{enumerate}[label=(\roman*)]
    \item \textbf{Energy bounded below:} For time-translations unitarily represented by
    \begin{align}
        U(t) = e^{-i H t} \, ,
    \end{align}
    there is a unique self-adjoint operator~$H$ whose spectrum is bounded from below.
    \item \textbf{Time-translation covariance:} For a system initially localized to~$V_0$, then
    \begin{align*}
        U(t) N(V_0) U^\dagger(t) = N(V_t) \, ,
    \end{align*}
    where~$V_t$ is the time-translated volume.
\end{enumerate}
The original formulation by~\textcite{hegerfeldt_remark_1974} additionally required covariance under spatial translations, which was essentially used to impose \ac{nw}'s~orthogonality postulate, albeit generalized for systems with compact support. Later, \textcite{hegerfeldt_causality_1998, hegerfeldt_instantaneous_1998} dropped this assumption---only requiring~(i) and~(ii)---but still making two further assumptions; adapting them from~\textcite{halvorson_no_2002}, they are:
\begin{enumerate}[label=(\roman*), resume]
    \item \textbf{Monotonicity:} If the system is localized in~$V$, then it is also localized in any larger volume~$V'$ containing~$V$.
    \item \textbf{No instantaneous wavepacket spreading:} If~$V_0$ is contained in~$V'_0$, and the boundaries of~$V_0$ and~$V'_0$ have a finite distance, then there exists an~$\epsilon > 0$ such that~$\braket{N(V_0)} \leq \braket{N(V'_t)}$ for all~$0 \leq t < \epsilon$.
\end{enumerate}
Accepting these conditions, Hegerfeldt showed that all four cannot simultaneously hold. Either the system must remain trivially localized, such that
\begin{align} \label{eq:first_case_hegerfeldt}
    \braket{\psi_t | N(V_0) | \psi_t} \equiv 1 \; \text{for all } t \, ,
\end{align}
or else 
\begin{align} \label{eq:second_case_hegerfeldt}
    \braket{\psi_t | N(V_0) | \psi_t} < 1 \; \text{for almost all } t \, .
\end{align}
For the trivial solution, the system is restricted to the initial volume~$V_0$ for all time, as would happen for a bound state in an external potential. In general, however, a system will not be confined to~$V_0$ for all time. In this case, the system immediately delocalizes and propagates superluminally, \textit{i.e.}, a free system initially localized at time~$t = 0$ can be localized only for that instant. 

One might worry that this superluminal propagation contradicts the microcausality of~\ac{rqt}, but as~\textcite{hegerfeldt_causality_1998, hegerfeldt_instantaneous_1998, hegerfeldt_particle_2001} has emphasized, this result crucially depends on the energy being bounded from below. There are no causality issues for hyperbolic wave equations, such as the Dirac and Klein-Gordon equations, since the spectrum of the Hamiltonian is unbounded. In this sense, Hegerfeldt's theorem also implicitly depends on the choice of unitary representation of the Poincar\'{e} group, \textit{viz.}~the hyperbolic and canonical representations. This connection between localization and dynamics is similarly elucidated by Malament's theorem.  


\subsubsection{Malament's theorem}


Motivated by Hegerfeldt's theorem and other aforementioned works, \textcite{malament_defense_1996} formulated a no-go theorem with the express intention of refuting the existence of localizable systems, \textit{viz.}~particles. As in Hegerfeldt's theorem, Malament assumed a projector~$P$ (notationally distinct, but functionally equivalent) that determines whether the system is localized in a volume~$V$. 
Four key postulates underlie Malament's theorem, with the first two essentially equivalent to those from Hegerfeldt's theorem, albeit with~(ii) additionally requiring covariance under spatial translations.
The third postulate is:
\begin{enumerate}[label=(\roman*), start=3]
    \item \textbf{Localizability:} If~$V$ and~$V'$ are disjoint volumes in~$\Sigma$, then
    \begin{align}
        P(V) P(V') = P(V') P(V) = 0 \, .
    \end{align}
\end{enumerate}
The intuition is that, if a particle is strictly localized in~$V$, then it definitely cannot be found in~$V'$ for disjoint volumes; for a formulation that better expresses this intuition, see~\textcite{busch_unsharp_1999}. Thus, condition~(iii) places the particle in one location, but says nothing about causality. The last postulate, which Malament originally introduced noncovariantly, is the only one that introduces causal structure:
\begin{enumerate}[label=(\roman*), resume]
    \item \textbf{Locality:} If~$V_t$ and~$V'_{t'}$ are spacelike-separated volumes defined with respect to a family of hypersurfaces~$\{ \Sigma_t \}$, then
    \begin{align}
        [P(V_t), P(V'_{t'})] = 0 \, .
    \end{align}
\end{enumerate}
However, given its essential connection with relativity, this postulate is most clearly stated covariantly. Associating~$P$ with spacetimes regions, it reads:
\begin{enumerate}[label=(\roman*)]
    \item[(iv$'$)] \textbf{Microcausality:} If~$\mathcal{O}$ and~$\mathcal{O}'$ are spacelike-separated spacetime regions, then
    \begin{align}
        [P(\mathcal{O}), P(\mathcal{O}')] = 0 \, ,
    \end{align}
\end{enumerate}
where~(iv) is recovered by restricting~$\mathcal{O}$ and~$\mathcal{O}'$ to respective spatial hypersurfaces~$\Sigma_t$ and~$\Sigma_{t'}$. Accepting these four postulates, alongside standard axioms of relativity and quantum theory, it follows that
\begin{align}
    P(V) \equiv 0 \; \text{for all } V \, .
\end{align}
But this implies that nothing can be found \textit{anywhere}, which certainly goes against commonsense intuition, not to mention all of physics. Thus, at least one condition must be dropped, and so long as we simultaneously impose (i)~positive energies, (ii)~translation covariance, and (iv$'$)~microcausality, then (iii)~localizability must go. However, while this result again demonstrates a conflict between localization and causality, it also reveals another connection---invoking conditions~(i) and~(ii) inseparably ties localization with dynamics, not just causality. 

\newpage

\subsection{Relativistic propagators}

In~\ac{nrqm}, the propagator is the Green function of the Schr\"odinger equation
\begin{align}
    G_{\mathrm{NR}}(\bm{x}, t; \, \bm{x}', t') := \braket{\bm{x} | e^{-i H (t - t')} | \bm{x}'} \, ,
\end{align}
which gives the transition amplitude between spatial states at different times, jointly characterizing both the localization and dynamics. 
As expected, this clear-cut picture in~\ac{nrqm} contrasts with the richer structure in~\ac{rqt}, where the two localization schemes lead to different dynamics---an aspect explored in quantum cosmology and the sum-over-histories approach starting in the mid-1980s. A seminal work by~\textcite{hartle_path_1986} investigated relativistic propagators in the canonical and path integral formulations of one-particle~\ac{rqm}, which was further developed by~\textcite{halliwell_sum-over-histories_1993}, and subsequently treated by various authors~\autocite{anderson_use_1994, halliwell_decoherent_2001, padmanabhan_obtaining_2018}. For the purposes of this discussion, let us compare the~\ac{nw} and causal propagators.


\subsubsection{Newton-Wigner propagator}

Recalling the positive-energy irreducible representation of Foldy's canonical form~\eqref{eq:canonical_form}, the dynamics of the \ac{nw}~scheme are given by the relativistic Schr\"{o}dinger equation
\begin{align*}
    i \partial_t \psi_{\mathrm{NW}}(x) = \sqrt{-\bm{\nabla}^2 + m^2} \, \psi_{\mathrm{NW}}(x) \, ,
\end{align*}
whose solutions are propagated according to
\begin{align}
    \psi_{\mathrm{NW}}(x') = \int d^3\bm{x} \, G_{\mathrm{NW}}(x'; \, x) \, \psi_{\mathrm{NW}}(x) \, ,
\end{align}
where~$G_{\mathrm{NW}}$ is the \ac{nw}~propagator, given by
\begin{align} \label{eq:nw_propagator}
    G_{\mathrm{NW}}(x; \, x') = \frac{1}{(2\pi)^3} \int d^3\bm{p} \, e^{-i p \cdot (x - x')} \, .
\end{align}
Since this propagator satisfies the Schr\"{o}dinger equation, it retains the properties from~\ac{nrqm}. Namely, the \ac{nw}~propagator satisfies the standard composition law
\begin{align} \label{eq:nw_propagator_transitivity}
    G_{\mathrm{NW}}(x''; \, x) = \int d^3\bm{x}' \, G_{\mathrm{NW}}(x''; \, x') \, G_{\mathrm{NW}}(x'; \, x) \, ,
\end{align}
and, as expected, it also satisfies orthogonality for equal times
\begin{align} \label{eq:nw_propagator_orthogonality}
    G_{\mathrm{NW}}(\bm{x}, t; \, \bm{x}', t') \Big|_{t = t'} = \delta^{(3)}(\bm{x} - \bm{x}') \, .
\end{align}
However, the \ac{nw}~propagator does not vanish outside the support of the light cone, displaying its characteristic superluminal propagation 
\begin{align}
    G_{\mathrm{NW}}(x'; \, x) \neq 0 \; \text{ for } (x' - x)^2 < 0 \, ,
\end{align}
and it is not Lorentz invariant, being dependent on the choice of hyperplane.



\subsubsection{Causal propagator}

In the hyperbolic representation, one has several definitions for the relativistic propagator. Following~\textcite{halliwell_sum-over-histories_1993} and~\textcite{fulling_aspects_1989}, these propagators can be obtained from the kernel of the \ac{kg}~equation, which satisfies
\begin{align}
    \left( \Box + m^2 \right) \mathcal{G}(x; y) = -\delta^{(4)}(x - y) \, ,
\end{align}
whose Fourier transform gives
\begin{align} \label{eq:klein_gordon_kernel}
    \mathcal{G}(x; y) = \frac{1}{(2\pi)^4} \int d^4\bm{p} \, \frac{e^{-i p \cdot (x - y)}}{p^2 - m^2} \, .
\end{align}
The different propagators are obtained depending on how one closes the contour around the poles at~$p_0 = \pm \sqrt{\bm{p}^2 + m^2}$. These different contour deformations give, for example, the Wightman functions, the Feynman propagator, and the causal propagator, but not the \ac{nw}~propagator. Of particular interest is the causal propagator, which propagates solutions of the \ac{kg}~equation, and is defined on globally hyperbolic spacetimes as
\begin{align}
    \phi(x') = -\int_\Sigma d\Sigma^\mu \, G(x'; \, x) \leftrightarrowoverset{\partial}_{\mu} \phi(x) \, ,
\end{align}
where~$\Sigma$ is a spacelike hypersurface and~$d\Sigma^\mu$ is a future-pointing spacetime volume element. Unlike the transitivity of the non-relativistic and \ac{nw}~propagators, the causal propagator satisfies the relativistic composition law
\begin{align}
    G(x''; \, x) = -\int_\Sigma d\Sigma^{\mu'} \, G(x''; \, x') \leftrightarrowoverset{\partial}_{\mu'} G(x'; \, x) \, ,
\end{align}
and does not satisfy equal-time orthogonality
\begin{align}
    G(\bm{x}, t; \, \bm{x}', t') \Big|_{t = t'} = 0 \, ,
\end{align}
but curiously \textit{does} display a kind of ``dynamical localization'' according to
\begin{align}
    \frac{\partial}{\partial t} G(\bm{x}, t; \, \bm{x}', t) \Big|_{t = t'} = -\delta^{(3)}(\bm{x} - \bm{x}') \, .
\end{align}
Additionally, as its name suggests, the causal propagator exhibits the expected causal behavior, namely having support within the light cone
\begin{align}
    G(x'; \, x) = 0 \; \text{ for } (x' - x)^2 < 0 \, ,
\end{align}
and being independent of a specified hyperplane, thus Lorentz invariant.

The mutually exclusive properties of each propagator concretely show the conflict expressed by Hegerfeldt's and Malament's theorems. Localizability is inseparable from causal dynamics, as is further exemplified by issues in the treatment of relativistic path integrals~\autocite{anderson_use_1994, halliwell_decoherent_2001, padmanabhan_obtaining_2018}. Having covered this background, let us summarize the localization problem and the historical positions taken in response.

\section{The localization problem} \label{sec:loc_problem}

The different frameworks constituting the localization problem naturally motivate different intuitions and positions. As illustrated thus far, conflicts exist between both localization and Lorentz invariance, and localization and relativistic causality. In response to these dichotomies, different camps have emerged to address the localization problem, which I represent as two alliterative characters: the ``Localizability Loyalist'' and the ``Causal Conformist.'' While the Loyalist insists that strict localization is a necessary feature of~\ac{rqt}, the Conformist maintains the obvious necessity of relativistic causality. Let us review the arguments of both sides. 

\subsection{Localizability Loyalism} \label{sec:loyalism}

Historically, the strongest proponents of localization have been the American physicists Irving E. Segal~(\citeyear{segal_quantum_1964, segal_representations_1967}) and Gordon N. Fleming~(\citeyear{fleming_covariant_1965, fleming_nonlocal_1965, fleming_manifestly_1966, fleming_lorentz_1988, fleming_hyperplane_1989, fleming_just_1996, fleming_strange_1999, fleming_reeh-schlieder_2000, fleming_observations_2003, fleming_observations_2004, fleming_bell_2016}), who I take to epitomize the Loyalist's position. Working within the algebraic approach to~\ac{rqft}, Segal developed a \ac{nw}-adjacent localization scheme, which he justified by straightforwardly asserting:
\begin{quote}
    Experimental apparatus is necessarily localized in space and time, and particles would not be observable without their localization in space at a particular time~\autocite[p.~139]{segal_quantum_1964}.
\end{quote}
Where, developing this intuition, Segal recasts~\ac{rqft} in a manifestly localized form much like Foldy's canonical representation. Accordingly,~\textcite{fleming_just_1996} provides a concise account of the Loyalist's position:\endnote{While Fleming sympathizes with this account, he does not defend strict localization~\autocite{fleming_observations_2004}, emphasizing instead a related notion of ``hyperplane dependence'' in \ac{rqm}.} 
\begin{quote}
    The Segal approach displays vividly the relationship between the locally covariant field structure [...]~and the non-locally covariant particle structure, which is closely related to state representations of the Newton-Wigner~(\citeyear{newton_localized_1949}), Foldy-Wouthuysen~(\citeyear{foldy_dirac_1950}) and Feschbach-Villars~(\citeyear{feshbach_elementary_1958}) form. It is the nature of this relationship which appears as an obstacle to the successful introduction of interactions and, as Saunders~[\citeyear{saunders_locality_1992}] points out, without the particle structure at hand we have no access to precise conceptions of spatio-temporally localized laboratory operations, the spatio-temporally localized field structures not being susceptible to an observable/probability distribution interpretation~\autocite[p.~12]{fleming_just_1996}.
\end{quote}
This spatio-temporally localized particle-like structure is the very thing reviewed throughout Secs.~\ref{sec:rel_loc} and~\ref{sec:loc_spin_half}, namely the \ac{nw}~localization scheme and its accompanying canonical representation. As I contend, it is this structure that provides the basis for a measurement framework constituted by observables, states, and probability assignments, which have been demonstrably supported by the following properties:
\begin{enumerate}[label=(\roman*)]
    \item self-adjoint position operator and, for spin-$1/2$ systems, a square-root Hamiltonian;
    \item energy bounded below with measurements preserving positivity of the energy;
    \item compatibility with an $L^2$-inner product for systems of arbitrary spin.
\end{enumerate}
Most significantly is, however, that one must fix a spacelike hypersurface when working with the \ac{nw}~localization scheme (for Minkowski spacetime, this corresponds to a hyperplane of simultaneity associated with an inertial observer). As noted in Sec.~\ref{sec:nw_loc}, a given state~$\ket{\Psi_{\mathrm{NW}}, \Sigma}$ must be defined relative to a spacelike hypersurface~$\Sigma$ at a fixed time. It is this property, above all others, that is conducive to the formal construction of the measurement framework.

These considerations are not specific to the \ac{nw}~localization scheme---several pioneers of~\ac{rqt}~\autocite{dirac_lagrangian_1933, tomonaga_relativistically_1946, schwinger_quantum_1948}, in attempting to define a relativistic state at a fixed time, recognized the need to introduce a spacelike hypersurface. Moreover, other recent approaches to measurement in~\ac{rqt} have also emphasized the significance of spacelike hypersurfaces~\autocite{fleming_manifestly_1966, bloch_relativistic_1967, aharonov_states_1980, aharonov_can_1981, aharonov_is_1984, fleming_lorentz_1986, fleming_lorentz_1988, fleming_hyperplane_1989}. However, such foliation-dependent state reduction is not uncontentious, and has been challenged by both~\textcite{maudlin_space-time_1996, maudlin_quantum_2011}, and~\textcite{myrvold_peaceful_2002} [but see the rejoinder by~\textcite{fleming_observations_2003}]. Nonetheless, within a relationalist metaphysics, it is natural to define observables only in relation to a given observer---here, with respect to a spacelike hypersurface at a fixed time. If an observer is not introduced in a hyperplane-dependent way, then a relationalist approach would otherwise have to model the measurement apparatus, which again raises the question of the apparatus' localization.

I hope my exposition above makes clear the Loyalist position. It is precisely the particle-like structure and the hyperplane dependence of the \ac{nw}~localization scheme that is essential for a measurement framework. However, as the Conformist points out, this very structure introduces serious problems: the \ac{nw}~localization scheme is not Lorentz invariant, and the time evolution of localized states conflicts with microcausality. How should the Loyalist respond? While I will not delve into specifics, Fleming has attempted to address these problems by introducing additional structure---namely, a radical form of hyperplane dependence where transformations act covariantly on both spacetime and the specified hyperplane, which as~\textcite{fleming_strange_1999} outline:
\begin{quote}
    [T]o specify the localized states, we need~[...]~three parameters in addition to those for the Minkowski coordinates of the localization - viz., parameters describing the hyperplane. Thus quantum localization takes place in a seven-dimensional manifold, rather than in four-dimensional Minkowski spacetime~\autocite[p.~131]{fleming_strange_1999}.
\end{quote}
Whatever one makes of this proposal, it \textit{still} does not resolve the unattractive dynamical properties of the \ac{nw}~localization scheme. The non-covariance of localized states is now accommodated by the Lorentz-covariant transformation of the hyperplane, but the time evolution of localized states still leads to superluminal propagation. 
Nonetheless,~\textcite{fleming_strange_1999} remain unperturbed:
\begin{quote}
    [We] believe that the superluminal propagation does not lead to causal contradictions, and is not in conflict with available empirical data: so that it should not be ruled out. We also believe that the other strange properties are merely unfamiliar novelties of [Lorentz-invariant quantum theory] \textit{which we must simply learn to accept}~\autocite[p.~109, emphasis added]{fleming_strange_1999}.
\end{quote}
The Conformist balks, as \textit{any} possibility of causal violations is completely unacceptable. Indeed, considering Hegerfeldt's and Malament's theorems, their concern is certainty justified. However strongly the Loyalist insists on the necessity of localization for a measurement framework, there is no obvious resolution to the problems posed by relativistic dynamics. Thus, the Conformist contently rests their case: give up localization for causality. But let us not be too hasty---we should cross-examine both sides before reaching a verdict. Let us consider the Conformist position.

\subsection{Causal Conformism} \label{sec:conformism}

Reviewing the arguments in favor of localization reveals few willing to take such a position. But while the Loyalists are few in number, there is no dearth of those defending causality. 
We have already seen several of its exponents in this review, namely~\textcite{malament_defense_1996}, and~\textcite{halvorson_no_2002}, but Conformism is implicit in practically every paper or textbook on~\ac{rqt}. After all, who would be so bold as to give up microcausality for a position operator? While the Loyalist might try to justify the superluminal propagation of the \ac{nw}~scheme, such assurances are met with persistent, reasonable skepticism. A provocative but compelling response on the causality issues, epitomizing the Conformist's position, was articulated by the American physicist Sidney Coleman~(\citeyear{coleman_lectures_2018}):
\begin{quote}
    Things are not so bad, however, as you would think. The particle doesn't have \textit{much} of a probability of traveling faster than light.~[...]~The chance that the particle is found outside of the forward light cone falls off exponentially as you get farther from the light cone. This makes it extremely unlikely that, for example, I could go back in time and convince my mother to have an abortion. But if it is at all possible, it is still unacceptable if you’re a purist. If you’re a purist, the~$X_i$~operator we have defined~[\textit{i.e.}, the \ac{nw}~position operator]~is absolutely rotten, no good, and to be rejected. If instead you’re a slob, it’s not so bad, because the amplitude of finding the particle outside of the forward light cone is rather small~\autocite[p.~14]{coleman_lectures_2018}.
\end{quote}
In response, the Loyalist might pose the following challenge: if localization is rejected, what of particle-like phenomena? The Conformist's ready answer is to weaken localization, but not discard it outright. This was already seen in the Philips scheme, wherein two Philips position states always have infinitesimal but nonvanishing overlap, even at opposite ends of the universe. As~\textcite{halvorson_no_2002} write:
\begin{quote}
    The argument for localizable particles appears to be very simple: Our experience shows us that objects (particles) occupy finite regions of space. But the reply to this argument is just as simple: These experiences are illusory! Although no object is strictly localized in a bounded region of space, an object can be well-enough localized to give the appearance to us (finite observers) that it is strictly localized. In fact, RQFT itself shows how the ``illusion'' of localizable particles can arise, and how talk about localizable particles can be a useful fiction~\autocite[p.~20]{halvorson_no_2002}.
\end{quote}
As previously discussed, a stronger line of argument by the Loyalist is to appeal to the necessity of localization for measurement, since the Conformist implicitly presupposes that a ready measurement framework---of observables, states, and probability assignments---already exists in~\ac{rqt}. This is assuredly untrue. As was stressed by~\textcite{fleming_strange_1999}, the localization problem extends to other structures pertaining to measurability, namely the choice of inner product and negative energies. For the latter, such issues persist even if one restricts to positive energies or promotes to a Fock space, since the energy density in~\ac{rqft} is generically nonpositive~\autocite{epstein_nonpositivity_1965}. Furthermore, the conflict between localization and causality is also present in $S$-matrix theory, regarding both the restriction to positive energies~\autocite{eden_problem_1965}, and the construction of relativistic quantum states~\autocite{blum_state_2017}. The Conformist is required not only to demonstrate consistence with particle-like phenomenology, but crucially to show the compatibility of a measurement framework with~\ac{rqt}, or at the very least how a measurement framework emerges in the low-energy regime.

Although \textcite{halvorson_no_2002} claim that delocalized observables can, in principle, be approximated by strictly localized counterparts, they do not offer a formal proposal. Within~\ac{rqm}, one attempt along these lines was provided by~\textcite{bracken_localizing_1999}, who recognized the difficulties in defining states and observables for spin-$1/2$ particles; they addressed these problems by generalizing states and observables, reconciling relativistic causality and weak localization. As~\textcite{melloy_generalized_2002} elaborated, these generalized states require additional structure, namely by redefining them as a class of sequences asymptotically approaching positive-energy states. While this particular construction can provide a measurement framework, it is unclear how general this solution is, and whether it generalizes for arbitrary spin.

However, even if the Conformist succeeds in reconciling causality and measurability, what then of the \ac{nw}~localization scheme? This question was raised by~\textcite{saunders_dissolution_1994}:
\begin{quote}
    NW-locality presents something of an enigma; for if it is relevant to particle positions, and if the latter are what are observed in the laboratory---at least in the low-energy limit of particle-detection phenomenology---then the events that we see do not appear to have a covariant description, in direct conflict with relativity. If, on the other hand, it is not relevant to particle positions, then one wonders what it can possibly signify, and why it is there. It is not an ad hoc construction~\autocite[p.~88]{saunders_dissolution_1994}.
\end{quote}
Thus, the burden of proof on the Conformist is not only to show that a comprehensive measurement framework can be constructed, consistent with causal dynamics, but \textit{also} to explain the role of the \ac{nw}~scheme. For if there is a ready-made formalism---providing both strict localization and a measurement framework---then the Conformist is obliged to relate their superior construction to the \ac{nw}~scheme, and justify how the latter emerges as a special case or otherwise.

\section{A situation within the foundations of physics} \label{sec:situation}

The Loyalist--Conformist debate seemingly leaves no room for compromise: either sacrifice microcausality for strict localization, or accept total delocalization to preserve it. If this dichotomy were an either-or choice, as it is always framed, then clearly the Conformist's position is favored. However, as the Loyalist has emphasized, giving up localizability is a serious concession, for it is a prerequisite for measurement. So understood, this problem is not merely a conflict between localization and causality---it is an antinomy between two equally necessary but mutually exclusive qualities.

To fully understand the localization problem, we must situate it more broadly within the foundations of physics. After all, its domain is not restricted to localizability but bears as much on measurement and causality. Moreover, this problem is not specific to quantum theory; it is already present in classical relativity~\autocite{pryce_mass-centre_1948}. Nor is this problem solely relativistic; it also arises in~\ac{nrqm}~\autocite{inonu_representations_1952}. But if issues are present in both~\ac{nrqm} and classical relativity, then what is the root of the problem? And if this problem is neither fundamentally relativistic nor quantum in origin, then might its resolution impact even classical nonrelativistic mechanics? While it is not possible to fully pursue these questions here, I will outline my own interpretation of the \ac{nw}~scheme---as regards measurability and temporal becoming---and reframe the Loyalist--Conformist debate so as to clarify the nature of this antinomy.


\subsection{Interpreting the Newton-Wigner localization scheme}


Recalling the postulates of the \ac{nw}~scheme, it is notable that no temporal extension is ever invoked---the localized states are defined to transform solely under the Euclidean group and are consequently restricted to a spacelike hyperplane. As the \ac{nw}~scheme lacks temporal extent, let us interpret it as characterizing an ``instant'' in time. Indeed, it aligns both with this concept and our commonsense notion of the ``present'', where we situate ourselves relative to things with definite spatial positions. 
These considerations should not be surprising, for localizability bears as much on the ``here'' as the ``now''. This connection was poetically articulated by~\textcite{stein_relativity_1991}:
\begin{quote}
    [T]he original meaning of the word ``present'' was not \textit{now}, but \textit{here-now} (\textit{i.e.}, ``nearby now''). Indeed, the original explicit meaning would seem rather to have been the spatial one (Latin \textit{praesens}, present participle of \textit{praeesse}---``to be in front of'', ``to be at hand'')---although, of course, with the temporal component understood (indeed, implied by the tense). That remains a current usage: When a soldier at roll call responds ``Present!'' upon hearing his name, he is not merely announcing that he still exists; he means that he is on the spot~\autocite[p.~159]{stein_relativity_1991}.
\end{quote}
However, the intuitive notion of a spatially unbounded present is rather alien from the perspective of modern physics and seems explicitly counter to a relativistically covariant description of spacetime. It is precisely this ``specious present'' that~\textcite{stein_relativity_1991} has objected to, seeking instead an alternative conception of a ``present'' compatible with relativity: a point in Minkowski spacetime. Yet such a point-like characterization loses the qualities of a spatially unbounded present, as would be provided by a spacelike hyperplane---a point of contention mentioned by subsequent authors~\autocite{clifton_definability_1995, dorato_becoming_1996, myrvold_relativistic_2003}.

This division---between space and spacetime---also occurs within~\ac{rqt}, albeit in another guise. The \ac{nw}~scheme and its canonical representation characterize an instantaneous, measurable mode of reality. 
It is precisely this instantaneity that is so essential, for one never perceives a system's temporal extension, but only ever its spatial manifestation.\endnote{Such instantaneous measurability should not be conflated with an act of measurement---the latter occurring over a period of time, but always as a series of instants. Temporal extension is never observed.} Yet such instantaneity is counter to a four-dimensional description of reality, which both has manifest temporal extension and furnishes a causal relation between points in Minkowski spacetime. 
In~\ac{rqt}, this causal mode of reality is formalized by the hyperbolic representation and its Lorentz-invariant ``localization'' scheme, but inadvertently sacrifices measurability---the relation between observer and observable---which is now consigned to the \ac{nw}~scheme.

While the Loyalist and Conformist have debated the significance of these two structures in~\ac{rqt}, this dichotomy also exists at the level of spacetime structure. The current discussion singles out two modes of reality: a three-dimensional, phenomenological instant versus a four-dimensional, causal spacetime. These two modes are likewise manifest within~\ac{nrqm}, delineating localizability and measurement from unitary dynamics, which becomes an essential distinction in~\ac{rqt}. Let us consider this duality further.

\subsection{Potentiality versus actuality}


Much has been written on the interpretations of quantum theory, particularly regarding measurement and entanglement, and their ontological implications. Recently, \textcite{fleming_actualization_1992} has discussed the neo-Aristotelian renditions of potentiality and actuality as a means to interpret quantum phenomena. As Fleming mentions, a notable defender of this metaphysics in quantum theory was the American physicist-philosopher Abner E. Shimony~(\citeyear{shimony_events_1986}), who characterized these two notions as follows:
\begin{quote}
    The combination of indefiniteness of value [\textit{viz.}, quantum superpositions] with definite probabilities of possible outcomes can be compactly referred to as \textit{potentiality}, a term suggested by~\textcite[p.~185]{heisenberg_physics_1958}. When a physical variable which initially is merely potential acquires a definite value, it can be said to be \textit{actualized}~\autocite[p.~142]{shimony_events_1986}.
\end{quote}
Insofar as potentiality and actualization have been applied to quantum theory, they are standardly invoked to describe the process of state reduction, where a superposition collapses to a definite measurement result. 
Shimony, however, questions this identification and offers instead the following:
\begin{quote}
    The conjecture is that \textit{the reduction of a superposition, however that is achieved, does not} ipso facto \textit{constitute the actualization of a physical variable of which the reduced state is an eigenstate}~\autocite[p.~153, emphasis in original]{shimony_events_1986}.
\end{quote}
By this, Shimony challenges a purely epistemic account of actualization, specifically addressing the realist presuppositions of~\textcite{einstein_can_1935}; by their account, a physical quantity is actualized if it can be predicted with certainty, \textit{i.e.}, a corresponding ``element of physical reality'' exists independently of observation. In contrast, \textcite{shimony_events_1986} proposes an alternative account of potentiality and actualization: 
\begin{quote}
    My conjecture, however, has a considerable metaphysical content, \textit{for it distinguishes between two modalities of reality}, and it modifies the criterion of Einstein~\textit{et al.} by abstaining from the identification of the `existence of an element of physical reality' with existence in the \textit{modality of actuality}~\autocite[p.~154, emphasis added]{shimony_events_1986}. 
\end{quote}
While the extent to which quantum theory demands such a metaphysical view is questionable, I intend to adapt Shimony’s conjecture to the present problem. Here, two distinct modalities seem apparent: one involving the definite, phenomenological reality of localized observables; the other, an indefinite, dynamical reality consistent with relativistic causality. Accordingly, state reduction should be distinguished from the metaphysical actualization occurring between these two modalities; as it pertains to~\ac{rqt}, one must first fix a spacelike hyperplane to define an ``instant'' upon which an idealized measurement can occur. If we accept this distinction---actuality versus potentiality---can any resolution to the antinomy between Loyalist and Conformist be found?

\subsection{Antinomy and synthesis}


Recognizing such dichotomies, let us note the different roles of physical operators decompose along the same lines. For instance, the Hamiltonian functions as both an observable for energy and a symmetry generator for time translations, while a relativistic spin operator similarly plays both an observable and dynamical role. 
This latter case was explicitly treated by~\textcite{terno_two_2003}, who remarked that these dual roles are typically not distinguished. 
For while the distinction is unneeded in~\ac{nrqm}, it becomes crucial for~\ac{rqt}: the \ac{nw}~scheme and canonical representation treat operators as observables, whereas the hyperbolic representation identifies them as symmetry generators.

As it pertains to the localization problem, the Loyalist and Conformist have been partly debating which of these two structures is fundamental. The Loyalist prioritizes a measurement framework of three-dimensional space, while the Conformist underscores the causal dynamics of four-dimensional spacetime, but each presupposes only one set of structures to be true. However, it is this very presupposition that must be challenged. If the formal structure of~\ac{rqt} is meant to describe physical reality, then there are two inequivalent descriptions. 
By all appearances, there is a duality between these two structures. Let us accept them as pertaining to dual modalities.

Considering two instants in time, there must be an intermediate time evolution relating such moments. But there cannot be causal evolution within an instant, for the Hamiltonian acts not as a symmetry generator but only as an observable, and attempts contrariwise lead to delocalization and acausality. Conversely, for any causal motion, there must be moments upon which phenomena can be measured, lest there be no connection to observable reality. The two modalities must be intertwined, wherein potentialities are actualized and actualities are potentialized. That is, this duality entails a modal involution, whereupon these observable and dynamical modes are interchanged. 

As for Hegerfeldt's and Malament's theorems, the nature of these dual modalities challenge the purportedly unassuming dynamical postulates, namely energy bounded from below and time translation covariance. But if a distinction is made between the observable and dynamical character of the Hamiltonian, then it is clear that these conditions do not have to be simultaneously satisfied. Indeed, the hyperbolic representation of~\ac{rqt} is plagued by the long-standing issue of negative energies---even when a Fock space structure is introduced and the fields are quantized, one still obtains negative energy densities in~\ac{rqft}~\autocite{epstein_nonpositivity_1965}. In actuality, at an instant in time, all observable energy densities are positive. In potentiality, where the Hamiltonian generates time translations, its spectrum is unbounded.


Simultaneously, the intertwining of these two modalities has rather dramatic implications for the ontology of~\ac{rqt}. For if a Fock space structure is insufficient to explain away such negative energies, then a common justification for taking quantum fields as the fundamental ontology is undermined. However, a particle ontology is also problematic inasmuch as it is incompatible with the hyperbolic representation, but this does not answer which of the two possibilities to prefer. The most conservative move is, perhaps, a return to the particle-wave duality of old quantum theory, albeit now understood in relation to the dual modalities. That the Loyalist and Conformist have debated the ontology of these structures as much as localizability is testament to the significance of this question. Whatever the case, a simple monistic ontology is disfavored by these considerations, although it is unclear to what extent our concepts must be revised.

\section{Concluding remarks} \label{sec:conclusion}

In this historical review, I have juxtaposed the \ac{nw}~and Lorentz-invariant schemes, and emphasized the inseparability of localization from causal dynamics, with the latter explicitly revealed by the Poincar\'e group's dual representations and their relativistic propagators. The conflict between localization and causality has produced two opposing camps---the Localizability Loyalist and the Causal Conformist---arguing for the necessity of one position. According to the received review, supporting the Conformist, localization must be sacrificed to ensure causal dynamics. In response, the Loyalist challenges that localization is a prerequisite for measurement, and this objection is not without merit. More fundamental issues are at hand.

One might still contend that there is no antinomy between localization and causality. But if it is simply a question of adopting a causal-dynamical reality---without localization---then what of the \ac{nw}~scheme? The ready presence of a measurement framework, comprising observables, states, and probability assignments, is an immediate feature of such structures. Even more, these problems are not specific to~\ac{rqt}, but also appear in both classical relativity and~\ac{nrqm}. If we take a moment to consider the matter afresh, then it certainly seems we are missing something quite fundamental. As I have argued, it is the very nature of reality---hitherto supposed as a single modality---that comes into question.

If we postulate not one, but two modes of reality---actuality and potentiality---then there is a ready answer to the antinomy before us, achieved by synthesis of the two modalities and their coincident formal structures. Of course, a duplication of reality seems an objectionable extravagance, but this is not what is being proposed. The dual modalities are meant to separate out two notions---observability and dynamics---that have hitherto been conjoined. This is not extravagance, but exigence. For in actuality, one has a ground upon which to stand, but no legs with which to move; while in potentiality, one has legs with which to move, but no ground upon which to stand. But irrespective of this view, the formal structures presented here demand explanation, and one that goes to the very foundations of physics---of where things are and how they move. 

\section*{Acknowledgments}

I would like to thank David Wallace, Doreen Fraser, Eirini Telali, Guilherme Franzmann, Ivan Romualdo de Oliveira, Jan Mandrysch, Jason Pye, Magdalena Zych, Maria Papageorgiou, Peter Evans, Tony Bracken, Viktoria Kabel, and Wayne Myrvold for comments and discussion. 

This research was supported by an Australian Government Research Training Program (RTP) Scholarship, the Australian Research Council Centre of Excellence for Engineered Quantum Systems (EQUS, CE170100009), and the Mitacs Globalink Research Award.

The author acknowledges the Turrbal and Jagera peoples, who are the Traditional Owners of the land on which the University of Queensland is situated.




\newpage

\printendnotes

\newpage

\printbibliography[heading=bibintoc]

\end{document}